\documentclass[prd,twocolumn,amsmath,amssymb,nofootinbib,floatfix,superscriptaddress]{revtex4-2}

\usepackage{times,graphics}
\usepackage[colorlinks]{hyperref}
\usepackage{makecell}
\usepackage{diagbox}

\usepackage{graphicx}
\usepackage{dcolumn}
\usepackage{bm}
\usepackage{natbib}
\usepackage{graphicx}
\usepackage{epsfig}
\usepackage{epstopdf}
\usepackage{multirow}
\usepackage{amsmath}
\usepackage{lipsum}
\usepackage{amssymb}
\usepackage{color}
\usepackage{dcolumn}
\usepackage{bm}
\usepackage{natbib}
\usepackage{tabu}
\usepackage{times,graphics}
\usepackage{makecell}
\usepackage{diagbox}
\usepackage{soul,xcolor}

\RequirePackage{mathptmx}      % use Times fonts if available on your TeX system
\RequirePackage{flushend}
\setstcolor{red}

\graphicspath{{./fig/}}

\begin{document}

\title{Gaussian discriminators between \texorpdfstring{$\Lambda$}{}CDM and wCDM cosmologies using expansion data}

\author{Ahmad Mehrabi }
\affiliation{Department of Physics, Bu-Ali Sina University, Hamedan	65178, 016016, Iran}
	
\author{Jackson Levi Said}
\affiliation{Institute of Space Sciences and Astronomy, University of Malta, Msida, Malta}
\affiliation{Department of Physics, University of Malta, Msida, Malta}

\begin{abstract}
The Gaussian linear model provides a unique way to obtain the posterior probability distribution as well as the Bayesian evidence analytically. Considering the expansion rate data, the Gaussian linear model can be applied for $\Lambda$CDM, wCDM and a non-flat $\Lambda$CDM. 
In this paper, we simulate the expansion data with various precision and obtain the Bayesian evidence, then it has been used to discriminate the models. The data uncertainty is in range $\sigma\in(0.5,10)\%$ and two different sampling rate have been considered. Our results indicate that it is possible to discriminate $w=-1.02$ (or $w=-0.98$) model from the $\Lambda$CDM $(w=-1)$ with $\sigma=0.5\%$ uncertainty in expansion rate data. Finally, we perform a parameters inference in both the MCMC and Gaussian linear model, using current available expansion rate data and compare the results.
\end{abstract}

\maketitle

\section{Introduction}

$\Lambda$CDM cosmology offers a simple and consistent concordance model which has agreed with observations for several decades \cite{Riess:1998cb,Perlmutter:1998np}. Separately, the $\Lambda$CDM model gives excellent agreement with cosmic microwave background data \cite{Planck:2018vyg,eBOSS:2020yzd}, as well as late time measurements of cosmic expansion \cite{Riess:2019cxk}. However, recent observations have suggested a growing cosmic tension in the reporting of the Hubble constant \cite{DiValentino:2020vhf,DiValentino:2020zio,Riess:2021jrx}, among other tensions. Together with the long standing consistency issues with the model \cite{Weinberg:1988cp}, this points to possible deviations from $\Lambda$CDM cosmology entering the observational regime. In this new context, it is imperative to understand the accuracy of observational data needed to discriminate between viable alternatives.

Cosmological tensions in the $\Lambda$CDM model has led to a re-evaluation of the foundations of the model such as the role of the cosmological principle \cite{Bengaly:2021wgc,Nadolny:2021hti}, as well as the nature of dark matter and dark energy (such as Refs.\cite{Copeland:2006wr,Benisty:2021gde,Benisty:2020otr}), and the fundamental description of gravity \cite{Clifton:2011jh,CANTATA:2021ktz,Bahamonde:2021gfp,AlvesBatista:2021gzc,Addazi:2021xuf}. One such alternative that has gained popularity in recent years is the wCDM model. This is composed of a dynamical equation of state for dark energy that varies across the cosmic history of the Universe. It remains an open question whether observational constraints can point to a varying equation of state in the near future. Thus, it is important to understand what precision would be needed to discriminate between these two models. 

Among all tensions and issues in the $\Lambda$CDM, the so called $H_0$ tension, is the most severe one. A Lot of efforts have been undertaken so far to tackle the problem without any reliable and satisfactory solution (to see more details refer to \cite{DiValentino:2021izs}). Notice that, along with all model modifications scenarios that have been considered so far, describing a data set in a model independent approach might be very useful in this case \cite{Vazirnia:2021xuu,Mehrabi:2020zau,Mehrabi:2021feg}.

For a typical cosmological setup, one naturally investigates a Markov chain Monte Carlo (MCMC) approach to infer parameter values using the latest observational data sets. However, given the plethora of cosmological models being proposed, this approach only gives the constrain on the free parameters and says nothing about model comparison. For this purpose, some statistical measures, under some simplifying assumptions, including the Akaike information criterion (AIC), the Bayesian information criterion (BIC) and Deviance information criterion (DIC) have been used for model selection. On the other hand the Bayesian evidence provides a robust and reliable measure for model selection. Unfortunately, computation of the quantity involves a high-dimension integration over the product of likelihood and prior, which is computationally very expensive. To overcome this, some numerical approaches like the nested sampling \cite{Handley:2015fda,nest-2020,Alsing:2021wef} and the Savage-Dickey density ratio \cite{Trotta:2005ar} have been developed. However, when the model is linear in its free parameters, the Gaussian linear model (GLM) provides an analytic solution for the Bayesian evidence. The formalism for a Gaussian or flat prior has been presented in \cite{Nesseris:2012cq}. Moreover, the Bayesian model selection has been utilized in \cite{Keeley:2021dmx,Koo:2021suo} to understand reliability of the Bayesian evidence. In this work, we consider three important cosmological models which are linear in their parameters and apply the GLM method to understand how precision of the expansion rate data affects the significance of the model discrimination through Bayesian evidence.

The structure of this paper is as follows:  In Sec.~\ref{sec:glm}, we give the basic formalism of the Gaussian linear model (GLM) and introduce the analytic formula to obtain the posterior distribution as well as the evidence. In Sec.~\ref{sec:sim}, the details of simulated data in our models are given. In addition, we present the results of applying the GLM on these data in the section. Then in Sec.~\ref{sec:obs}, we describe current available Hubble data from different observations and perform an MCMC parameter inference to obtain the best value of parameters as well as their uncertainties. We also apply the GLM method to the observational data and compare the results with those of the MCMC. Finally, we conclude and discuss the main points of our finding in section Sec.~\ref{conclude}.

\section{Gaussian linear model}\label{sec:glm}
In this section, we briefly provide the basic formalism of the GLM. In this scenario, a database is modeled by a function which is linear in its parameters 
\begin{equation}\label{eq:model}
f(x,\theta) = \sum \theta_j X^j(x),
\end{equation}   
where $X(x)$ is an arbitrary function of x and $\theta_j$s are the free parameters. Notice that the base functions can very well be a non-linear function of x. Assuming a $n_{obs}$ dimension database as $(x_i,y_i,\tau_i)$, the likelihood function is given by
 \begin{equation}\label{eq:likelihood}
 p(y|\theta) = \mathcal{L}_0\exp[-\frac{1}{2}(\theta-\theta_0)^tL(\theta-\theta_0)]
 \end{equation}
 where,
\begin{equation}\label{eq:likelihood-nor}
    \mathcal{L}_0= \frac{1}{(2\pi)^{n_{obs}/2}\Pi\tau_i} \exp[-\frac{1}{2}(b-A\theta_0)^tL(b-A\theta_0)],
\end{equation}
 and 
\begin{equation}\label{eq:L-theta_0}
    F_{ij}=X^j(x_i)~~,~~A= \frac{F_{ij}}{\tau_i}~~,~~b=\frac{y_i}{\tau_i}~~,~~L=A^tA~~,~~\theta_0=L^{-1}A^tb.
\end{equation}
Here the maximum likelihood occurs at $\theta_0$  and $L$ denotes the likelihood Fisher matrix. 
 
In order to perform the Bayesian parameter inference, we need to define a prior on the free parameters. We consider a Gaussian prior as
\begin{equation}\label{eq:prior}
p(\theta) = \frac{|\Sigma_{pri}|^{-1/2}}{(2\pi)^{n_{par}/2}}\exp[-\frac{1}{2}(\theta-\theta_{pri})^t\Sigma_{pri}^{-1}(\theta-\theta_{pri})],
 \end{equation}
 where $\theta_{pri}$ ($\Sigma_{pri}$) is the mean (covariance matrix) of the prior and $n_{par}$ denotes the number of free parameters. Using the Bays theorem, the posterior distribution is proportional to
  \begin{equation}\label{eq:pos}
  p(\theta|y) \propto \exp[-\frac{1}{2}(\theta-\theta_{pos})^t\Sigma_{pos}^{-1}(\theta-\theta_{pos})],
  \end{equation}
  where $$\Sigma_{pos}^{-1} = \Sigma_{pri}^{-1} + L $$ and $$\theta_{pos}=(\Sigma_{pri}^{-1} + L)^{-1}(\Sigma_{pri}^{-1}\theta_{pri}+L\theta_0)$$.
  
In addition to the Bayesian parameter inference, the GLM provides the Baysian evidence which is a key quantity  in model comparison. The Bayesian evidence includes an integration over all parameters space and is given by
  \begin{equation}\label{eq:evid_gen}
p(y) = \int d\theta p(\theta)p(\theta|y).
\end{equation}
Since both the prior and likelihood are a multivariate Gaussian in the GLM, the integral has an analytical solution.
The Bayesian evidence in the GLM is given by, 
  \begin{equation}\label{eq:evid}
  p(y) = \mathcal{L}_0|\Sigma_{pri}|^{-1/2}|\Sigma_{pos}|^{1/2}\exp(D),
  \end{equation}
  where 
  \begin{eqnarray}\label{eq:D}
  D &=& \frac{1}{2}[(\theta_{pri}^t\Sigma_{pri}^{-1} + \theta_{0}^tL)(\Sigma_{pri}^{-1}+L)^{-1}(\Sigma_{pri}^{-1}\theta_{pri}
   +L\theta_0) \nonumber\\ &-& (\theta_{pri}^t\Sigma_{pri}^{-1}\theta_{pri} + \theta_{0}^tL\theta_{0})],
  \end{eqnarray} 
 and $|\Sigma|$ denotes the determinant of $\Sigma$. In Addition, as we mentioned before, it is possible to have a solution in the case of the flat priors. To see more details refer to \cite{Nesseris:2012cq}.
 
When comparing two models $\mathcal{M}_0$ and $\mathcal{M}_1$, using the Bayes theorem, it is straightforward to obtain,
  \begin{equation}\label{eq:bay-fac}
\frac{p(\mathcal{M}_0|d)}{p(\mathcal{M}_1|d)} = B_{01}\frac{p(\mathcal{M}_0)}{p(\mathcal{M}_1)},
\end{equation}
where $B_{01}$ is the Bayes factor. Usually, the prior on the models $p(\mathcal{M})$ is taken to be flat and so the Bayes factor is the key quantity in Bayesian model comparison. The value of Bayes factor should be interpreted by an empirically calibrated scale to compare given models. The Jeffreys’ \cite{Jeffreys61} and the Kass-Raftery scales \cite{kass-raf} provides two well-known scales to interpret the Bayes factor. These two scales are presented in Tab.~\ref{tab:jef} and Tab.~\ref{tab:kas}.

\begin{table}
	\centering
	\begin{tabular}{|c|c|}
		\hline
			$|\ln B_{01}|$  &  Strength of evidence \\ \hline
			$<1$ &  Inconclusive \\ \hline
			$1.$ &  Weak evidence\\  \hline
			$2.5$ & Moderate evidence \\ \hline
			$5$ &  Strong evidence\\  \hline
	\end{tabular}
     \caption{The Jeffreys’ scales for interpreting the Bayes factor.}
	\label{tab:jef}
\end{table}

\begin{table}
	\centering
	\begin{tabular}{|c|c|}
		\hline
			$|\ln B_{01}|$  &  Strength of evidence \\ \hline
			0 to 1 &  Inconclusive \\ \hline
			1 to 3 &  Positive evidence\\  \hline
			3 to 5 &  Strong evidence \\ \hline
			$>$ 5 &  Very strong evidence\\  \hline
	\end{tabular}
     \caption{The Kass-Raftery scales for interpreting the Bayes factor.}
	\label{tab:kas}
\end{table} 

Notice that these evidences are in favor of the model with larger evidence.

\section{Simulated data and results}\label{sec:sim}
In order to apply the GLM in the cosmological context, we should have a linear model. Considering the $\Lambda$CDM model, the Hubble parameter as a function of redshift  can be written as 
\begin{eqnarray}\label{eq:LCDM}
    H^2(z) &=& 100^2 [\Omega_{m}h^2(1+z)^3 + h^2-\Omega_{m}h^2]\\ \nonumber
    &=&100^2 [\Omega_{m}h^2((1+z)^3-1) + h^2],
\end{eqnarray}
where $H_0=100h$ is the current expansion rate of the universe and $\Omega_{m}$ is the matter density parameter.
Interestingly, the second format is linear in its parameters and can be seen as a GLM with 
$$X^1(z) = (1+z)^3-1~~,~~X^2(z)=1$$ and free parameters $$\theta_1 = \Omega_{m}h^2~~,~~\theta_2=h^2 $$.

In addition to the $\Lambda$CDM, the wCDM and non-flat $\Lambda$CDM (N$\Lambda$CDM ) also can be written as a GLM model,
\begin{equation}\label{eq:nf_LCDM}
H^2(z) = 100^2 [\Omega_{m}h^2((1+z)^3-(1+z)^{3(1+w)}) + h^2(1+z)^{3(1+w)}],
\end{equation} 
for the wCDM and 
\begin{equation}\label{eq:wCDM}
H^2(z) = 100^2 [\Omega_{m}h^2((1+z)^3-1) + \Omega_{k}h^2((1+z)^2-1) + h^2],
\end{equation} 
for the N$\Lambda$CDM. In this case, the $\Omega_{k}$ is the curvature density parameter. Since these models are GLM, it is easy to find the value of parameters which maximize the likelihood, the mean and covariance of posterior distribution as well as the Bayesian evidence. 
  
To obtain the posterior distributions as well as the evidence, we need to define a prior on the free parameters. To avoid any possible prior bias, we consider a Gaussian wide prior on the free parameters. We use a multivariate Gaussian with mean and covariance matrix as
\begin{equation}\label{eq:pri_mean}
\theta_{pri} = (\Omega_{m}h^2=0.13,\Omega_{k}h^2=0.,h^2=0.45)
\end{equation} 

\begin{equation}\label{eq:pri_cov}
	\begin{bmatrix} 
0.05 & 0 & 0 \\
0 & 0.1 & 0\\
0 & 0 & 0.1 \\
\end{bmatrix}
\end{equation}
where the first, second and third row present covariance of the $\Omega_{m}h^2$, $\Omega_{k}h^2$ and $h^2$ respectively. We examine different mean and covarince matrix to check the robustness of our results. As long as the priors are wide enough, our results are the same and there is no prior bias.

\subsection{\texorpdfstring{$\Lambda$}{}CDM and wCDM}   
\begin{figure}[h]
	\centering
	\includegraphics[width=0.5\textwidth]{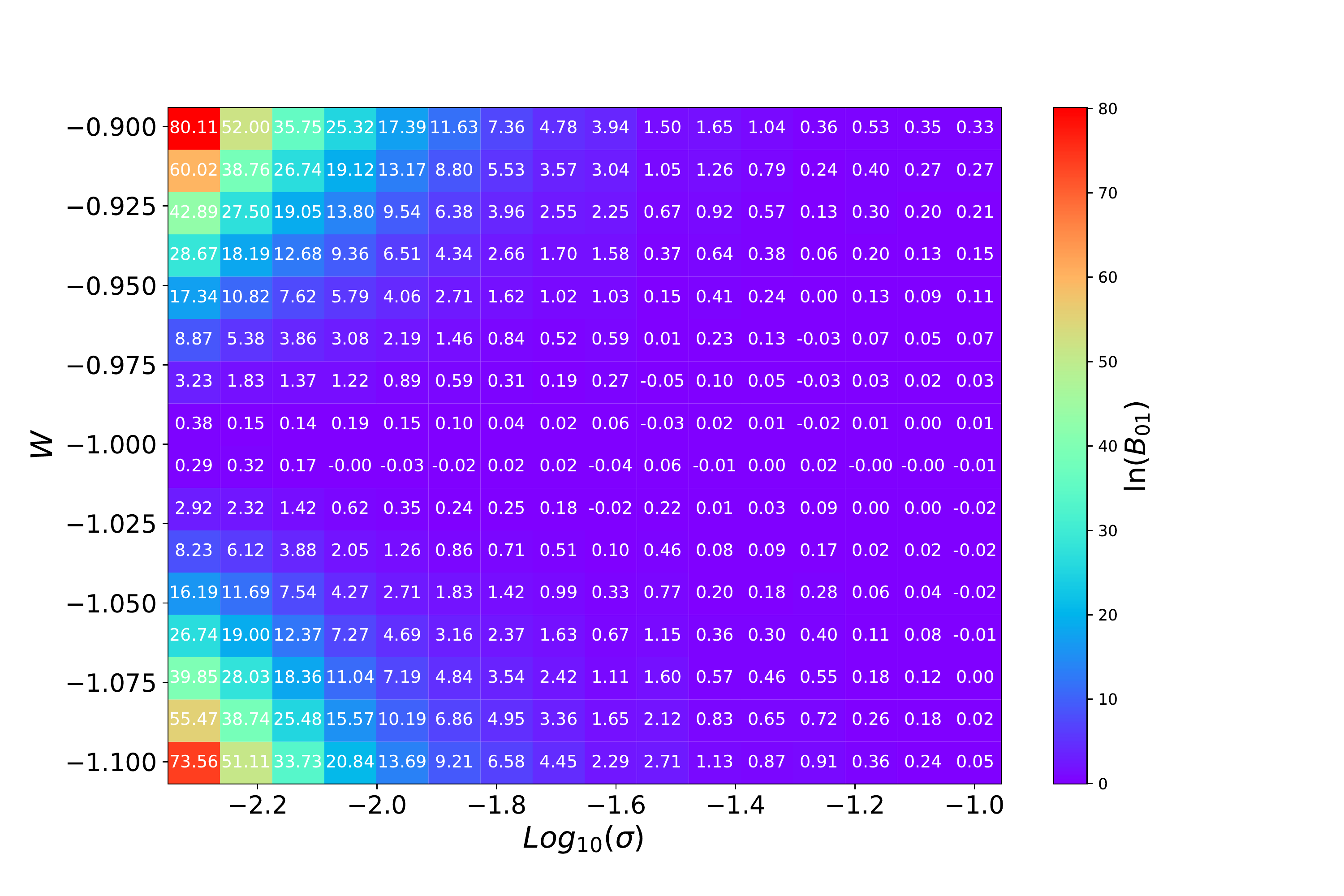}	
	\caption{The mean Bayes factor for the $\Lambda$CDM and wCDM using the first sampling strategy.  }
	\label{fig:lw1}
\end{figure}

\begin{figure}[h]
	\centering
	\includegraphics[width=0.5\textwidth]{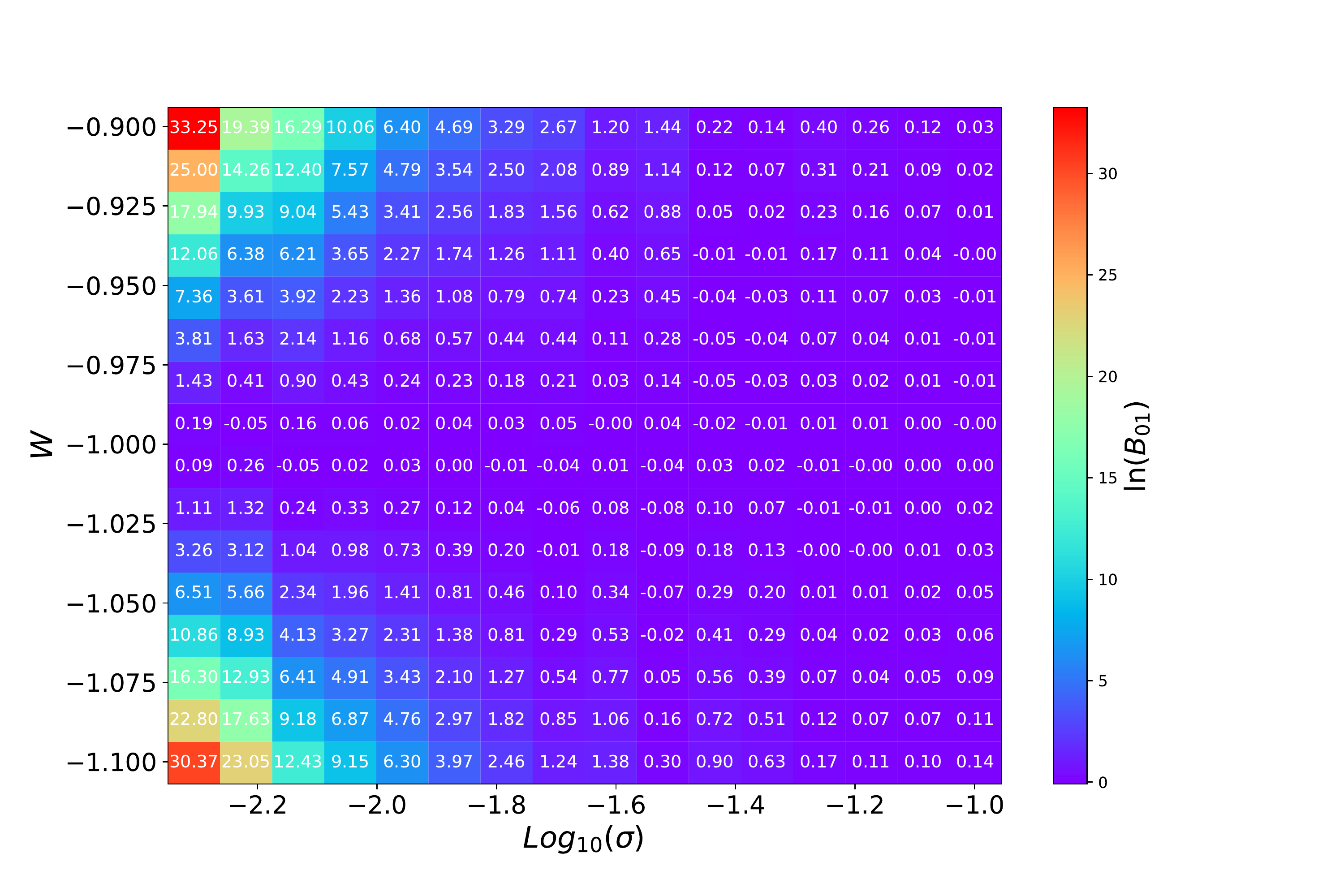}	
	\caption{The mean Bayes factor for the $\Lambda$CDM and wCDM using the second sampling strategy. }
	\label{fig:lw2}
\end{figure}

In order to realize how precision and sampling rate of an expansion rate database affects the model comparison,  we simulate the expansion rate data with different precision and use the GLM method to perform a model comparison. To compare the $\Lambda$CDM and wCDM,  two sampling rates have been considered. In the first case, we simulate 100 data points in range $z\in(0,3)$ with uncertainty $\sigma\in(0.5,10\%)$ using the $\Lambda$CDM model. For the second sampling strategy, we simulate 50 data points in the redshift range $z\in(0,2)$ and a similar uncertainty range as the first one. Now, we consider the simulated data and compute the evidence for both $\Lambda$CDM and wCDM. In our analysis, the EoS, is selected in a range $(w\in -0.9,-1.1)$. Moreover, we consider a Gaussian distribution for the uncertainty and so the data points are randomly generated at each simulation. Taking this into account, we have a distribution of the Bayes factor. In order to consider such a statistical fluctuation, we have generated 40 simulated data sets and compute the mean of Bayes factor. Notice that, we examine other values for the number of data sets and the results are quite the same by considering more data sets. The mean Bays factor $\ln B_{01}=\ln B_{\Lambda}-\ln B_{w}$ for two strategies, have been shown in Figs (\ref{fig:lw1}) and (\ref{fig:lw2}) respectively. The value of Bays factor for each cell is shown as a numerical value on the cell.     

From Fig.(\ref{fig:lw1}), it is clear that discriminating $\Lambda$CDM from wCDM with uncertainty larger than $10^{(-1.4)}\sim 4\%$ for a wide range of EoS is almost impossible. With $\sim 3\%$ uncertainty, we see a strong evidence only for $w\sim-0.9$ or $w\sim-1.1$, which is already disfavored by other observations. On the other hand, with $\sigma \leq 1\%$, the chance of discriminating increases significantly. For example with $\sigma=0.5\%$, a $2\%$ deviation in the EoS of dark energy ($w=1.02$ or $w=0.98$) could be detected with a strong evidence. Notice that, as we mentioned above, the base model for simulated data is the $\Lambda$CDM and the Bayes factor is computed for the $\Lambda$CDM and wCDM. Contrary, if we consider the wCDM as the base model for simulated data and compute the Bayes factor as $\ln B_{01}=\ln B_{w}-\ln B_{\Lambda}$, the results are the same. In this case, a positive Bayes factor indicates more evidence in favor of the wCDM.

In addition to the precision of each data point, the sampling rate of a database affects the model comparison. For a less cadence database, the results are presented in (\ref{fig:lw2}). Overall, the results are the same as the first sampling strategy but strength of the Bayes factor decreases. For example the extreme case in the first sampling strategy is $\ln B_{01}\sim 80$ for $\sigma=0.5\%$ and $w=-0.9$, while considering the second strategy, the number decreases to $\ln B_{01}\sim 33$ which is around $60\%$ less than the former.

\subsection{\texorpdfstring{$\Lambda$}{}CDM and the non-flat \texorpdfstring{$\Lambda$}{}CDM}   
\begin{figure}[h]
	\centering
	\includegraphics[width=0.5\textwidth]{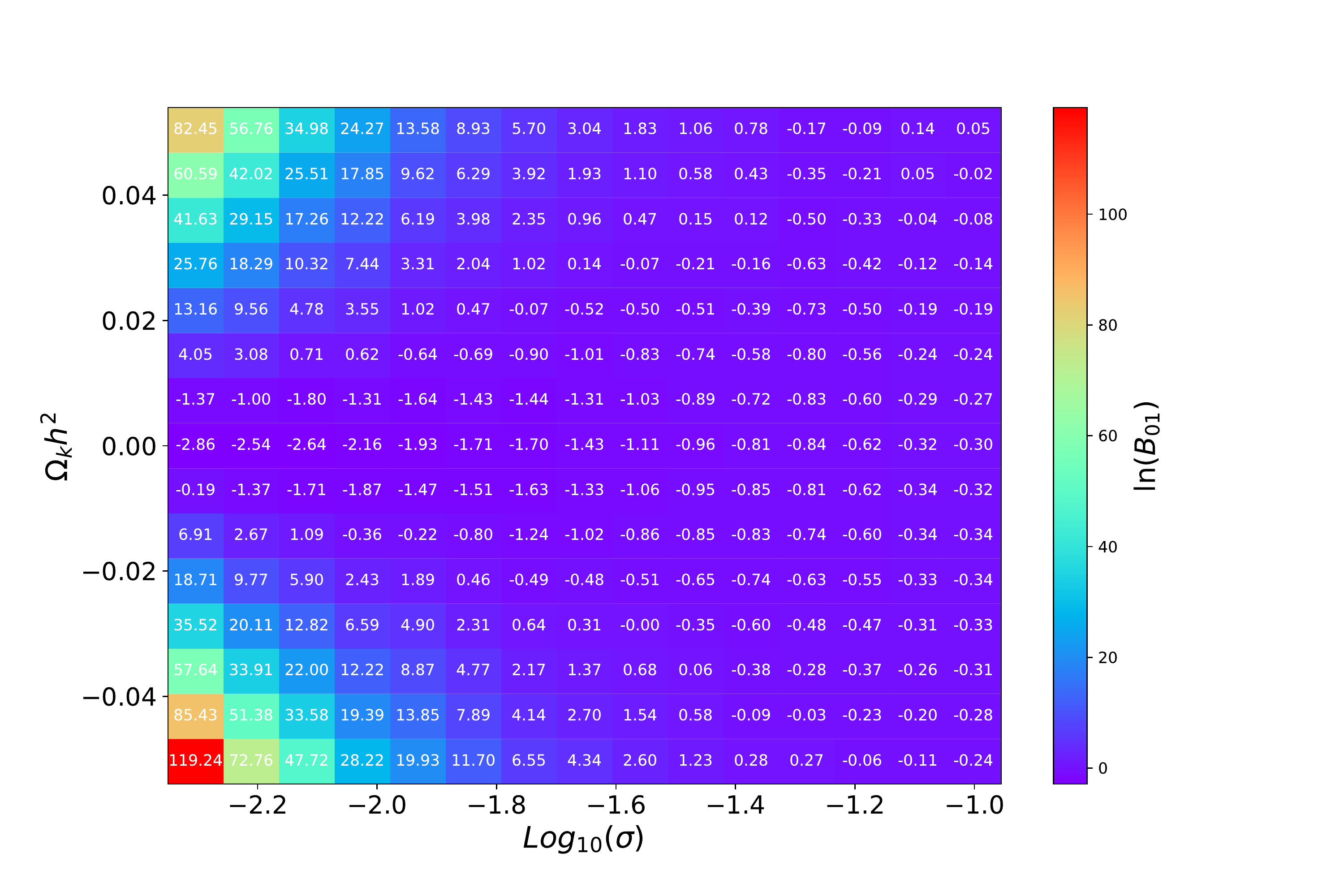}	
	\caption{The mean Bayes factor for the $\Lambda$CDM and N$\Lambda$CDM using the first sampling strategy.}
	\label{fig:lk1}
\end{figure}

\begin{figure}[h]
	\centering
	\includegraphics[width=0.5\textwidth]{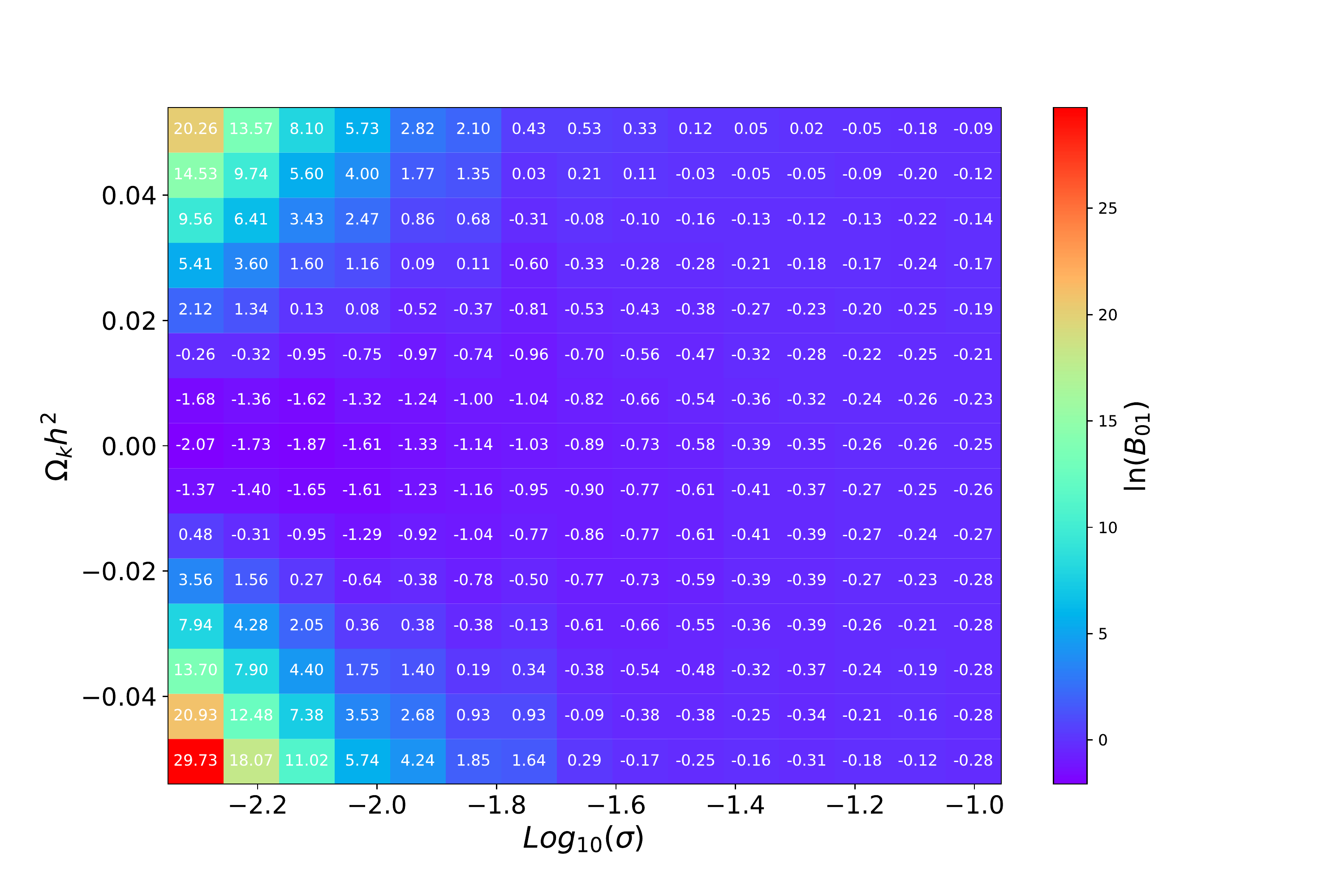}	
	\caption{The mean Bayes factor for the $\Lambda$CDM and N$\Lambda$CDM using the second sampling strategy. }
	\label{fig:lk2}
\end{figure}

 As we mentioned above, the  N$\Lambda$CDM can also be written in the form of a GLM. In this case, we follow similar strategies as previous one. The parameter $\Omega_{k}h^2$ is selected in the range $\Omega_{k}h^2\in(-0.05,0.05)$ to simulate data points. In this case, we simulate the Hubble data using N$\Lambda$CDM and compute the evidence in both N$\Lambda$CDM and $\Lambda$CDM. In Figs (\ref{fig:lk1}) and (\ref{fig:lk2}) the mean Bayes factor (averaging over 40 databases) $\ln B_{01}=\ln B_{N} - \ln B_{\Lambda} $ have been shown for two sampling strategies. Here $B_{N}$ indicates the evidence of the N$\Lambda$CDM model.   
 
Our results indicate that with $\sigma>3\%$ there is no chance to discriminate a flat and non flat $\Lambda$CDM models. The evidence become more significant at both positive and negative curvature with smaller uncertainties. We see very strong evidence for $\Omega_{k}h^2\sim 0.02$ with $\sigma<1\%$ which is much higher for larger and smaller values of $\Omega_{k}h^2$. Furthermore, interestingly, we see a negative evidence for a high accuracy data $\sigma<1\%$ when $\Omega_{k}h^2\sim 0$. This is due to the Occam’s razor effect which favor a simpler model. The effect indicates that an extra free parameter, not being constrained significantly with the data, makes the evidence smaller compare to the model without that free parameter. In these cases, the Bayes factor favor the simpler model which is in our case the flat $\Lambda$CDM. Of course the constrain for $\Omega_{k}h^2$ is significantly improved for larger and smaller value of $\Omega_{k}h^2$ so the evidence become positive and the more complex model is favored.    

The results for the second sampling strategy have been shown in (\ref{fig:lk2}). As the previous case, the strength of the evidence decreases for a less cadence sampling rate. In this case, the extreme case decreases around $75\%$ in the second sampling rate.

\section{Evidence for Current available Hubble data}\label{sec:obs}

In this section, we apply the GLM approach on the current observational data. The expansion rate database in our analysis is a combination of the Hubble parameter measurement from cosmic chronometers and the BAO measurements. The database has been collected in \cite{Farooq:2016zwm}. In addition, the local $H_0$ measurement (the SHOES data) \cite{Riess:2019cxk} has been added to the database. Considering the $\Lambda$CDM, wCDM and the non-flat $\Lambda$CDM, the results of MCMC analysis have been shown in the Table (\ref{tab:res}). To perform the MCMC analysis, we use the public python package {\it{pymc3}} \cite{pymc}. In this analysis, we consider the wide Gaussian prior on the free parameters introduced in section \ref{sec:sim}. The $1\sigma$ uncertainty of each parameters has been shown along with its mean value. These values are estimated from a sample of parameters generated in the MCMC algorithm.     

Now, we use the database and apply the GLM formalism to obtain the MLE, mean and covariance of posterior as well as the evidence. 
The results have been presented in Tab.(\ref{tab:res-glm}). As it is clear, the results from the GLM are quit in agreement with those of the MCMC. Notice that, in the case of GLM, we have an analytic posterior distribution for the free parameters and compute the mean and $1\sigma$ uncertainty directly from the distributions.         

\begin{table}
	\centering
	\begin{tabular}{|c|c|c|c|}
		\hline
			Model/Parameters  &  $h^2$ & $\Omega_{m}h^2$   & $\Omega_{k}h^2$ \\ \hline
			$\Lambda$CDM & $0.518\pm0.015$& $0.123\pm0.005$& -- \\ \hline
			N$\Lambda$CDM & $0.526\pm0.018$ & $0.135\pm0.016$& $-0.034\pm0.044$ \\ \hline
	\end{tabular}
     \caption{The mean and uncertainty of parameters in  $\Lambda$CDM and  N$\Lambda$CDM using MCMC method.}
	\label{tab:res}
\end{table}

\begin{table}
	\centering
	\begin{tabular}{|c|c|c|c|}
		\hline
		Model/Parameters  &  $h^2$ & $\Omega_{m}h^2$   & $\Omega_{k}h^2$ \\ \hline
		$\Lambda$CDM & $0.517\pm0.015$& $0.121\pm0.005$& -- \\ \hline
		N$\Lambda$CDM & $0.526\pm0.018$ & $0.134\pm0.016$& $-0.038\pm0.044$ \\ \hline
	\end{tabular}
	\caption{The mean and uncertainty of parameters in  $\Lambda$CDM and  N$\Lambda$CDM using the GLM method.}
	\label{tab:res-glm}
\end{table} 

In our analysis, we find $\ln B_{01}=\ln B_{\Lambda}-\ln B_{N}=0.43$ which indicates an inconclusive evidence for considered models. In fact, this result was expected because of the low precision observational data points. Notice that the most data points from cosmic chronometer have uncertainty larger than $10\%$ but uncertainty of the expansion rate data from BAO is less than $10\%$ and the uncertainty of the most precise one is $\sigma\sim3.5\%$.

\section{Conclusion}\label{conclude}

The landscape of cosmological models has drastically increased in recent years with the combined open problems of cosmological tensions and the internal consistency issues of gravitational models. In this work, we explore the GLM in the context of three cosmological models, namely $\Lambda$CDM, non-flat $\Lambda$CDM and the wCDM, in order to explore the question of precision requirements for specific data sets to discriminate these models. We wish to assess the data set precision needed to differentiate between each pair of these cosmological models.

Vanilla $\Lambda$CDM is our base model for considering any modification to the concordance model. Here, we compare $\Lambda$CDM together with wCDM where the equation of state parameter is allowed to be any value. These two model are central to modern cosmology and have Friedmann equations represented by Eq.~\eqref{eq:LCDM} and Eq.~\eqref{eq:wCDM} respectively. In addition, we also consider $\Lambda$CDM with a possible non-flat component in Eq.~\eqref{eq:nf_LCDM}. There have been recent suggestions in the literature that such a scenario may be preferable in the context of recent reporting by the Planck collaboration data \cite{DiValentino:2019qzk}.

The GLM approach presented here takes Friedmann equation components, and through a calculation resulting in the Bayes factor, can determine whether enough precision is present to differentiate between pairs of models. In Figs.~(\ref{fig:lw1},\ref{fig:lw2}) this is done for the $\Lambda$CDM and wCDM models where two sampling strategies are shown with very consistent results. Here, we find that indeed for an equation of state parameter that veers away from the $\Lambda$CDM value, the approach indicates a high confidence for differentiating these models. Specifically, we see a strong evidence in favor of the $\Lambda$CDM when the data simulated with uncertainty $\sigma\sim0.5$ and the rival model is the wCDM with $w=-1.02$ or $w=-0.98$. Moreover, we show how different sampling rates affects the Bayes factor. The pattern of the results are almost the same for our two sampling strategies but the value of the Bayes factor is smaller in the case of the less cadence sampling. 

In Figs.~(\ref{fig:lk1},\ref{fig:lk2}) we repeat the analysis for the $\Lambda$CDM and a non-flat $\Lambda$CDM setting. Our results indicate that it is impossible to discriminate these models with $\sigma>3\%$ even if the $\Omega_{k}h^2$ is in the range $(-0.05,0.05)$. The strength of the evidence become much more larger for $\Omega_{k}h^2$ not close to zero with uncertainty $\sigma<3\%$. In addition, we see clear evidence of Occam’s razor effect when the curvature density is close to zero. To make this procedure as transparent as possible, we are making the code for this analysis available for others to use and improve upon\footnote{\url{https://github.com/Ahmadmehrabi/GLM}}.

%As future work, we hope to extend the applicability of this approach to nonlinear scenarios where the Friedmann equations in the method tolerates more sophisticated models with more nuances.

%=================================================================

\bibliographystyle{apsrev4-1}
\bibliography{ref}

\begin{thebibliography}{35}%
\makeatletter
\providecommand \@ifxundefined [1]{%
 \@ifx{#1\undefined}
}%
\providecommand \@ifnum [1]{%
 \ifnum #1\expandafter \@firstoftwo
 \else \expandafter \@secondoftwo
 \fi
}%
\providecommand \@ifx [1]{%
 \ifx #1\expandafter \@firstoftwo
 \else \expandafter \@secondoftwo
 \fi
}%
\providecommand \natexlab [1]{#1}%
\providecommand \enquote  [1]{``#1''}%
\providecommand \bibnamefont  [1]{#1}%
\providecommand \bibfnamefont [1]{#1}%
\providecommand \citenamefont [1]{#1}%
\providecommand \href@noop [0]{\@secondoftwo}%
\providecommand \href [0]{\begingroup \@sanitize@url \@href}%
\providecommand \@href[1]{\@@startlink{#1}\@@href}%
\providecommand \@@href[1]{\endgroup#1\@@endlink}%
\providecommand \@sanitize@url [0]{\catcode `\\12\catcode `\$12\catcode
  `\&12\catcode `\#12\catcode `\^12\catcode `\_12\catcode `\%12\relax}%
\providecommand \@@startlink[1]{}%
\providecommand \@@endlink[0]{}%
\providecommand \url  [0]{\begingroup\@sanitize@url \@url }%
\providecommand \@url [1]{\endgroup\@href {#1}{\urlprefix }}%
\providecommand \urlprefix  [0]{URL }%
\providecommand \Eprint [0]{\href }%
\providecommand \doibase [0]{http://dx.doi.org/}%
\providecommand \selectlanguage [0]{\@gobble}%
\providecommand \bibinfo  [0]{\@secondoftwo}%
\providecommand \bibfield  [0]{\@secondoftwo}%
\providecommand \translation [1]{[#1]}%
\providecommand \BibitemOpen [0]{}%
\providecommand \bibitemStop [0]{}%
\providecommand \bibitemNoStop [0]{.\EOS\space}%
\providecommand \EOS [0]{\spacefactor3000\relax}%
\providecommand \BibitemShut  [1]{\csname bibitem#1\endcsname}%
\let\auto@bib@innerbib\@empty
%</preamble>
\bibitem [{\citenamefont {Riess}\ \emph {et~al.}(1998)\citenamefont {Riess}
  \emph {et~al.}}]{Riess:1998cb}%
  \BibitemOpen
  \bibfield  {author} {\bibinfo {author} {\bibfnamefont {A.~G.}\ \bibnamefont
  {Riess}} \emph {et~al.} (\bibinfo {collaboration} {Supernova Search Team}),\
  }\href {\doibase 10.1086/300499} {\bibfield  {journal} {\bibinfo  {journal}
  {Astron.J.}\ }\textbf {\bibinfo {volume} {116}},\ \bibinfo {pages} {1009}
  (\bibinfo {year} {1998})},\ \Eprint {http://arxiv.org/abs/astro-ph/9805201}
  {arXiv:astro-ph/9805201 [astro-ph]} \BibitemShut {NoStop}%
%%CITATION = ASTRO-PH/9805201;%%
\bibitem [{\citenamefont {Perlmutter}\ \emph {et~al.}(1999)\citenamefont
  {Perlmutter} \emph {et~al.}}]{Perlmutter:1998np}%
  \BibitemOpen
  \bibfield  {author} {\bibinfo {author} {\bibfnamefont {S.}~\bibnamefont
  {Perlmutter}} \emph {et~al.} (\bibinfo {collaboration} {Supernova Cosmology
  Project}),\ }\href {\doibase 10.1086/307221} {\bibfield  {journal} {\bibinfo
  {journal} {Astrophys.J.}\ }\textbf {\bibinfo {volume} {517}},\ \bibinfo
  {pages} {565} (\bibinfo {year} {1999})},\ \Eprint
  {http://arxiv.org/abs/astro-ph/9812133} {arXiv:astro-ph/9812133 [astro-ph]}
  \BibitemShut {NoStop}%
%%CITATION = ASTRO-PH/9812133;%%
\bibitem [{\citenamefont {Aghanim}\ \emph {et~al.}(2020)\citenamefont {Aghanim}
  \emph {et~al.}}]{Planck:2018vyg}%
  \BibitemOpen
  \bibfield  {author} {\bibinfo {author} {\bibfnamefont {N.}~\bibnamefont
  {Aghanim}} \emph {et~al.} (\bibinfo {collaboration} {Planck}),\ }\href
  {\doibase 10.1051/0004-6361/201833910} {\bibfield  {journal} {\bibinfo
  {journal} {Astron. Astrophys.}\ }\textbf {\bibinfo {volume} {641}},\ \bibinfo
  {pages} {A6} (\bibinfo {year} {2020})},\ \Eprint
  {http://arxiv.org/abs/1807.06209} {arXiv:1807.06209 [astro-ph.CO]}
  \BibitemShut {NoStop}%
\bibitem [{\citenamefont {Alam}\ \emph {et~al.}(2021)\citenamefont {Alam} \emph
  {et~al.}}]{eBOSS:2020yzd}%
  \BibitemOpen
  \bibfield  {author} {\bibinfo {author} {\bibfnamefont {S.}~\bibnamefont
  {Alam}} \emph {et~al.} (\bibinfo {collaboration} {eBOSS}),\ }\href {\doibase
  10.1103/PhysRevD.103.083533} {\bibfield  {journal} {\bibinfo  {journal}
  {Phys. Rev. D}\ }\textbf {\bibinfo {volume} {103}},\ \bibinfo {pages}
  {083533} (\bibinfo {year} {2021})},\ \Eprint
  {http://arxiv.org/abs/2007.08991} {arXiv:2007.08991 [astro-ph.CO]}
  \BibitemShut {NoStop}%
\bibitem [{\citenamefont {Riess}\ \emph {et~al.}(2019)\citenamefont {Riess},
  \citenamefont {Casertano}, \citenamefont {Yuan}, \citenamefont {Macri},\ and\
  \citenamefont {Scolnic}}]{Riess:2019cxk}%
  \BibitemOpen
  \bibfield  {author} {\bibinfo {author} {\bibfnamefont {A.~G.}\ \bibnamefont
  {Riess}}, \bibinfo {author} {\bibfnamefont {S.}~\bibnamefont {Casertano}},
  \bibinfo {author} {\bibfnamefont {W.}~\bibnamefont {Yuan}}, \bibinfo {author}
  {\bibfnamefont {L.~M.}\ \bibnamefont {Macri}}, \ and\ \bibinfo {author}
  {\bibfnamefont {D.}~\bibnamefont {Scolnic}},\ }\href {\doibase
  10.3847/1538-4357/ab1422} {\bibfield  {journal} {\bibinfo  {journal}
  {Astrophys. J.}\ }\textbf {\bibinfo {volume} {876}},\ \bibinfo {pages} {85}
  (\bibinfo {year} {2019})},\ \Eprint {http://arxiv.org/abs/1903.07603}
  {arXiv:1903.07603 [astro-ph.CO]} \BibitemShut {NoStop}%
%%CITATION = ARXIV:1903.07603;%%
\bibitem [{\citenamefont {Di~Valentino}\ \emph
  {et~al.}(2021{\natexlab{a}})\citenamefont {Di~Valentino} \emph
  {et~al.}}]{DiValentino:2020vhf}%
  \BibitemOpen
  \bibfield  {author} {\bibinfo {author} {\bibfnamefont {E.}~\bibnamefont
  {Di~Valentino}} \emph {et~al.},\ }\href {\doibase
  10.1016/j.astropartphys.2021.102606} {\bibfield  {journal} {\bibinfo
  {journal} {Astropart. Phys.}\ }\textbf {\bibinfo {volume} {131}},\ \bibinfo
  {pages} {102606} (\bibinfo {year} {2021}{\natexlab{a}})},\ \Eprint
  {http://arxiv.org/abs/2008.11283} {arXiv:2008.11283 [astro-ph.CO]}
  \BibitemShut {NoStop}%
\bibitem [{\citenamefont {Di~Valentino}\ \emph
  {et~al.}(2021{\natexlab{b}})\citenamefont {Di~Valentino} \emph
  {et~al.}}]{DiValentino:2020zio}%
  \BibitemOpen
  \bibfield  {author} {\bibinfo {author} {\bibfnamefont {E.}~\bibnamefont
  {Di~Valentino}} \emph {et~al.},\ }\href {\doibase
  10.1016/j.astropartphys.2021.102605} {\bibfield  {journal} {\bibinfo
  {journal} {Astropart. Phys.}\ }\textbf {\bibinfo {volume} {131}},\ \bibinfo
  {pages} {102605} (\bibinfo {year} {2021}{\natexlab{b}})},\ \Eprint
  {http://arxiv.org/abs/2008.11284} {arXiv:2008.11284 [astro-ph.CO]}
  \BibitemShut {NoStop}%
\bibitem [{\citenamefont {Riess}\ \emph {et~al.}(2021)\citenamefont {Riess}
  \emph {et~al.}}]{Riess:2021jrx}%
  \BibitemOpen
  \bibfield  {author} {\bibinfo {author} {\bibfnamefont {A.~G.}\ \bibnamefont
  {Riess}} \emph {et~al.},\ }\href@noop {} {\  (\bibinfo {year} {2021})},\
  \Eprint {http://arxiv.org/abs/2112.04510} {arXiv:2112.04510 [astro-ph.CO]}
  \BibitemShut {NoStop}%
\bibitem [{\citenamefont {Weinberg}(1989)}]{Weinberg:1988cp}%
  \BibitemOpen
  \bibfield  {author} {\bibinfo {author} {\bibfnamefont {S.}~\bibnamefont
  {Weinberg}},\ }\href {\doibase 10.1103/RevModPhys.61.1} {\bibfield  {journal}
  {\bibinfo  {journal} {Rev. Mod. Phys.}\ }\textbf {\bibinfo {volume} {61}},\
  \bibinfo {pages} {1} (\bibinfo {year} {1989})}\BibitemShut {NoStop}%
\bibitem [{\citenamefont {Bengaly}(2021)}]{Bengaly:2021wgc}%
  \BibitemOpen
  \bibfield  {author} {\bibinfo {author} {\bibfnamefont {C.}~\bibnamefont
  {Bengaly}},\ }\href@noop {} {\  (\bibinfo {year} {2021})},\ \Eprint
  {http://arxiv.org/abs/2111.06869} {arXiv:2111.06869 [gr-qc]} \BibitemShut
  {NoStop}%
\bibitem [{\citenamefont {Nadolny}\ \emph {et~al.}(2021)\citenamefont
  {Nadolny}, \citenamefont {Durrer}, \citenamefont {Kunz},\ and\ \citenamefont
  {Padmanabhan}}]{Nadolny:2021hti}%
  \BibitemOpen
  \bibfield  {author} {\bibinfo {author} {\bibfnamefont {T.}~\bibnamefont
  {Nadolny}}, \bibinfo {author} {\bibfnamefont {R.}~\bibnamefont {Durrer}},
  \bibinfo {author} {\bibfnamefont {M.}~\bibnamefont {Kunz}}, \ and\ \bibinfo
  {author} {\bibfnamefont {H.}~\bibnamefont {Padmanabhan}},\ }\href {\doibase
  10.1088/1475-7516/2021/11/009} {\bibfield  {journal} {\bibinfo  {journal}
  {JCAP}\ }\textbf {\bibinfo {volume} {11}},\ \bibinfo {pages} {009} (\bibinfo
  {year} {2021})},\ \Eprint {http://arxiv.org/abs/2106.05284} {arXiv:2106.05284
  [astro-ph.CO]} \BibitemShut {NoStop}%
\bibitem [{\citenamefont {Copeland}\ \emph {et~al.}(2006)\citenamefont
  {Copeland}, \citenamefont {Sami},\ and\ \citenamefont
  {Tsujikawa}}]{Copeland:2006wr}%
  \BibitemOpen
  \bibfield  {author} {\bibinfo {author} {\bibfnamefont {E.~J.}\ \bibnamefont
  {Copeland}}, \bibinfo {author} {\bibfnamefont {M.}~\bibnamefont {Sami}}, \
  and\ \bibinfo {author} {\bibfnamefont {S.}~\bibnamefont {Tsujikawa}},\ }\href
  {\doibase 10.1142/S021827180600942X} {\bibfield  {journal} {\bibinfo
  {journal} {Int. J. Mod. Phys. D}\ }\textbf {\bibinfo {volume} {15}},\
  \bibinfo {pages} {1753} (\bibinfo {year} {2006})},\ \Eprint
  {http://arxiv.org/abs/hep-th/0603057} {arXiv:hep-th/0603057} \BibitemShut
  {NoStop}%
\bibitem [{\citenamefont {Benisty}\ and\ \citenamefont
  {Staicova}(2021{\natexlab{a}})}]{Benisty:2021gde}%
  \BibitemOpen
  \bibfield  {author} {\bibinfo {author} {\bibfnamefont {D.}~\bibnamefont
  {Benisty}}\ and\ \bibinfo {author} {\bibfnamefont {D.}~\bibnamefont
  {Staicova}},\ }\href@noop {} {\  (\bibinfo {year} {2021}{\natexlab{a}})},\
  \Eprint {http://arxiv.org/abs/2107.14129} {arXiv:2107.14129 [astro-ph.CO]}
  \BibitemShut {NoStop}%
\bibitem [{\citenamefont {Benisty}\ and\ \citenamefont
  {Staicova}(2021{\natexlab{b}})}]{Benisty:2020otr}%
  \BibitemOpen
  \bibfield  {author} {\bibinfo {author} {\bibfnamefont {D.}~\bibnamefont
  {Benisty}}\ and\ \bibinfo {author} {\bibfnamefont {D.}~\bibnamefont
  {Staicova}},\ }\href {\doibase 10.1051/0004-6361/202039502} {\bibfield
  {journal} {\bibinfo  {journal} {Astron. Astrophys.}\ }\textbf {\bibinfo
  {volume} {647}},\ \bibinfo {pages} {A38} (\bibinfo {year}
  {2021}{\natexlab{b}})},\ \Eprint {http://arxiv.org/abs/2009.10701}
  {arXiv:2009.10701 [astro-ph.CO]} \BibitemShut {NoStop}%
\bibitem [{\citenamefont {Clifton}\ \emph {et~al.}(2012)\citenamefont
  {Clifton}, \citenamefont {Ferreira}, \citenamefont {Padilla},\ and\
  \citenamefont {Skordis}}]{Clifton:2011jh}%
  \BibitemOpen
  \bibfield  {author} {\bibinfo {author} {\bibfnamefont {T.}~\bibnamefont
  {Clifton}}, \bibinfo {author} {\bibfnamefont {P.~G.}\ \bibnamefont
  {Ferreira}}, \bibinfo {author} {\bibfnamefont {A.}~\bibnamefont {Padilla}}, \
  and\ \bibinfo {author} {\bibfnamefont {C.}~\bibnamefont {Skordis}},\ }\href
  {\doibase 10.1016/j.physrep.2012.01.001} {\bibfield  {journal} {\bibinfo
  {journal} {Phys. Rept.}\ }\textbf {\bibinfo {volume} {513}},\ \bibinfo
  {pages} {1} (\bibinfo {year} {2012})},\ \Eprint
  {http://arxiv.org/abs/1106.2476} {arXiv:1106.2476 [astro-ph.CO]} \BibitemShut
  {NoStop}%
%%CITATION = ARXIV:1106.2476;%%
\bibitem [{\citenamefont {Saridakis}\ \emph {et~al.}(2021)\citenamefont
  {Saridakis} \emph {et~al.}}]{CANTATA:2021ktz}%
  \BibitemOpen
  \bibfield  {author} {\bibinfo {author} {\bibfnamefont {E.~N.}\ \bibnamefont
  {Saridakis}} \emph {et~al.} (\bibinfo {collaboration} {CANTATA}),\
  }\href@noop {} {\  (\bibinfo {year} {2021})},\ \Eprint
  {http://arxiv.org/abs/2105.12582} {arXiv:2105.12582 [gr-qc]} \BibitemShut
  {NoStop}%
\bibitem [{\citenamefont {Bahamonde}\ \emph {et~al.}(2021)\citenamefont
  {Bahamonde}, \citenamefont {Dialektopoulos}, \citenamefont
  {Escamilla-Rivera}, \citenamefont {Farrugia}, \citenamefont {Gakis},
  \citenamefont {Hendry}, \citenamefont {Hohmann}, \citenamefont {Said},
  \citenamefont {Mifsud},\ and\ \citenamefont
  {Di~Valentino}}]{Bahamonde:2021gfp}%
  \BibitemOpen
  \bibfield  {author} {\bibinfo {author} {\bibfnamefont {S.}~\bibnamefont
  {Bahamonde}}, \bibinfo {author} {\bibfnamefont {K.~F.}\ \bibnamefont
  {Dialektopoulos}}, \bibinfo {author} {\bibfnamefont {C.}~\bibnamefont
  {Escamilla-Rivera}}, \bibinfo {author} {\bibfnamefont {G.}~\bibnamefont
  {Farrugia}}, \bibinfo {author} {\bibfnamefont {V.}~\bibnamefont {Gakis}},
  \bibinfo {author} {\bibfnamefont {M.}~\bibnamefont {Hendry}}, \bibinfo
  {author} {\bibfnamefont {M.}~\bibnamefont {Hohmann}}, \bibinfo {author}
  {\bibfnamefont {J.~L.}\ \bibnamefont {Said}}, \bibinfo {author}
  {\bibfnamefont {J.}~\bibnamefont {Mifsud}}, \ and\ \bibinfo {author}
  {\bibfnamefont {E.}~\bibnamefont {Di~Valentino}},\ }\href@noop {} {\
  (\bibinfo {year} {2021})},\ \Eprint {http://arxiv.org/abs/2106.13793}
  {arXiv:2106.13793 [gr-qc]} \BibitemShut {NoStop}%
\bibitem [{\citenamefont {Alves~Batista}\ \emph {et~al.}(2021)\citenamefont
  {Alves~Batista} \emph {et~al.}}]{AlvesBatista:2021gzc}%
  \BibitemOpen
  \bibfield  {author} {\bibinfo {author} {\bibfnamefont {R.}~\bibnamefont
  {Alves~Batista}} \emph {et~al.},\ }\href@noop {} {\  (\bibinfo {year}
  {2021})},\ \Eprint {http://arxiv.org/abs/2110.10074} {arXiv:2110.10074
  [astro-ph.HE]} \BibitemShut {NoStop}%
\bibitem [{\citenamefont {Addazi}\ \emph {et~al.}(2021)\citenamefont {Addazi}
  \emph {et~al.}}]{Addazi:2021xuf}%
  \BibitemOpen
  \bibfield  {author} {\bibinfo {author} {\bibfnamefont {A.}~\bibnamefont
  {Addazi}} \emph {et~al.},\ }\href@noop {} {\  (\bibinfo {year} {2021})},\
  \Eprint {http://arxiv.org/abs/2111.05659} {arXiv:2111.05659 [hep-ph]}
  \BibitemShut {NoStop}%
\bibitem [{\citenamefont {Di~Valentino}\ \emph
  {et~al.}(2021{\natexlab{c}})\citenamefont {Di~Valentino}, \citenamefont
  {Mena}, \citenamefont {Pan}, \citenamefont {Visinelli}, \citenamefont {Yang},
  \citenamefont {Melchiorri}, \citenamefont {Mota}, \citenamefont {Riess},\
  and\ \citenamefont {Silk}}]{DiValentino:2021izs}%
  \BibitemOpen
  \bibfield  {author} {\bibinfo {author} {\bibfnamefont {E.}~\bibnamefont
  {Di~Valentino}}, \bibinfo {author} {\bibfnamefont {O.}~\bibnamefont {Mena}},
  \bibinfo {author} {\bibfnamefont {S.}~\bibnamefont {Pan}}, \bibinfo {author}
  {\bibfnamefont {L.}~\bibnamefont {Visinelli}}, \bibinfo {author}
  {\bibfnamefont {W.}~\bibnamefont {Yang}}, \bibinfo {author} {\bibfnamefont
  {A.}~\bibnamefont {Melchiorri}}, \bibinfo {author} {\bibfnamefont {D.~F.}\
  \bibnamefont {Mota}}, \bibinfo {author} {\bibfnamefont {A.~G.}\ \bibnamefont
  {Riess}}, \ and\ \bibinfo {author} {\bibfnamefont {J.}~\bibnamefont {Silk}},\
  }\href {\doibase 10.1088/1361-6382/ac086d} {\bibfield  {journal} {\bibinfo
  {journal} {Class. Quant. Grav.}\ }\textbf {\bibinfo {volume} {38}},\ \bibinfo
  {pages} {153001} (\bibinfo {year} {2021}{\natexlab{c}})},\ \Eprint
  {http://arxiv.org/abs/2103.01183} {arXiv:2103.01183 [astro-ph.CO]}
  \BibitemShut {NoStop}%
\bibitem [{\citenamefont {Vazirnia}\ and\ \citenamefont
  {Mehrabi}(2021)}]{Vazirnia:2021xuu}%
  \BibitemOpen
  \bibfield  {author} {\bibinfo {author} {\bibfnamefont {M.}~\bibnamefont
  {Vazirnia}}\ and\ \bibinfo {author} {\bibfnamefont {A.}~\bibnamefont
  {Mehrabi}},\ }\href {\doibase 10.1103/PhysRevD.104.123530} {\bibfield
  {journal} {\bibinfo  {journal} {Phys. Rev. D}\ }\textbf {\bibinfo {volume}
  {104}},\ \bibinfo {pages} {123530} (\bibinfo {year} {2021})},\ \Eprint
  {http://arxiv.org/abs/2107.11539} {arXiv:2107.11539 [astro-ph.CO]}
  \BibitemShut {NoStop}%
\bibitem [{\citenamefont {Mehrabi}\ and\ \citenamefont
  {Basilakos}(2020)}]{Mehrabi:2020zau}%
  \BibitemOpen
  \bibfield  {author} {\bibinfo {author} {\bibfnamefont {A.}~\bibnamefont
  {Mehrabi}}\ and\ \bibinfo {author} {\bibfnamefont {S.}~\bibnamefont
  {Basilakos}},\ }\href {\doibase 10.1140/epjc/s10052-020-8221-2} {\bibfield
  {journal} {\bibinfo  {journal} {Eur. Phys. J. C}\ }\textbf {\bibinfo {volume}
  {80}},\ \bibinfo {pages} {632} (\bibinfo {year} {2020})},\ \Eprint
  {http://arxiv.org/abs/2002.12577} {arXiv:2002.12577 [astro-ph.CO]}
  \BibitemShut {NoStop}%
\bibitem [{\citenamefont {Mehrabi}\ \emph {et~al.}(2021)\citenamefont {Mehrabi}
  \emph {et~al.}}]{Mehrabi:2021feg}%
  \BibitemOpen
  \bibfield  {author} {\bibinfo {author} {\bibfnamefont {A.}~\bibnamefont
  {Mehrabi}} \emph {et~al.},\ }\href {\doibase 10.1093/mnras/stab2915}
  {\bibfield  {journal} {\bibinfo  {journal} {Mon. Not. Roy. Astron. Soc.}\
  }\textbf {\bibinfo {volume} {509}},\ \bibinfo {pages} {224} (\bibinfo {year}
  {2021})},\ \Eprint {http://arxiv.org/abs/2107.08820} {arXiv:2107.08820
  [astro-ph.CO]} \BibitemShut {NoStop}%
\bibitem [{\citenamefont {Handley}\ \emph {et~al.}(2015)\citenamefont
  {Handley}, \citenamefont {Hobson},\ and\ \citenamefont
  {Lasenby}}]{Handley:2015fda}%
  \BibitemOpen
  \bibfield  {author} {\bibinfo {author} {\bibfnamefont {W.~J.}\ \bibnamefont
  {Handley}}, \bibinfo {author} {\bibfnamefont {M.~P.}\ \bibnamefont {Hobson}},
  \ and\ \bibinfo {author} {\bibfnamefont {A.~N.}\ \bibnamefont {Lasenby}},\
  }\href {\doibase 10.1093/mnrasl/slv047} {\bibfield  {journal} {\bibinfo
  {journal} {Mon. Not. Roy. Astron. Soc.}\ }\textbf {\bibinfo {volume} {450}},\
  \bibinfo {pages} {L61} (\bibinfo {year} {2015})},\ \Eprint
  {http://arxiv.org/abs/1502.01856} {arXiv:1502.01856 [astro-ph.CO]}
  \BibitemShut {NoStop}%
\bibitem [{\citenamefont {Speagle}(2020)}]{nest-2020}%
  \BibitemOpen
  \bibfield  {author} {\bibinfo {author} {\bibfnamefont {J.~S.}\ \bibnamefont
  {Speagle}},\ }\href {\doibase 10.1093/mnras/staa278} {\bibfield  {journal}
  {\bibinfo  {journal} {Monthly Notices of the Royal Astronomical Society}\
  }\textbf {\bibinfo {volume} {493}},\ \bibinfo {pages} {3132–3158} (\bibinfo
  {year} {2020})}\BibitemShut {NoStop}%
\bibitem [{\citenamefont {Alsing}\ and\ \citenamefont
  {Handley}(2021)}]{Alsing:2021wef}%
  \BibitemOpen
  \bibfield  {author} {\bibinfo {author} {\bibfnamefont {J.}~\bibnamefont
  {Alsing}}\ and\ \bibinfo {author} {\bibfnamefont {W.}~\bibnamefont
  {Handley}},\ }\href {\doibase 10.1093/mnrasl/slab057} {\bibfield  {journal}
  {\bibinfo  {journal} {Mon. Not. Roy. Astron. Soc.}\ }\textbf {\bibinfo
  {volume} {505}},\ \bibinfo {pages} {L95} (\bibinfo {year} {2021})},\ \Eprint
  {http://arxiv.org/abs/2102.12478} {arXiv:2102.12478 [astro-ph.IM]}
  \BibitemShut {NoStop}%
\bibitem [{\citenamefont {Trotta}(2007)}]{Trotta:2005ar}%
  \BibitemOpen
  \bibfield  {author} {\bibinfo {author} {\bibfnamefont {R.}~\bibnamefont
  {Trotta}},\ }\href {\doibase 10.1111/j.1365-2966.2007.11738.x} {\bibfield
  {journal} {\bibinfo  {journal} {Mon. Not. Roy. Astron. Soc.}\ }\textbf
  {\bibinfo {volume} {378}},\ \bibinfo {pages} {72} (\bibinfo {year} {2007})},\
  \Eprint {http://arxiv.org/abs/astro-ph/0504022} {arXiv:astro-ph/0504022}
  \BibitemShut {NoStop}%
\bibitem [{\citenamefont {Nesseris}\ and\ \citenamefont
  {Garcia-Bellido}(2013)}]{Nesseris:2012cq}%
  \BibitemOpen
  \bibfield  {author} {\bibinfo {author} {\bibfnamefont {S.}~\bibnamefont
  {Nesseris}}\ and\ \bibinfo {author} {\bibfnamefont {J.}~\bibnamefont
  {Garcia-Bellido}},\ }\href {\doibase 10.1088/1475-7516/2013/08/036}
  {\bibfield  {journal} {\bibinfo  {journal} {JCAP}\ }\textbf {\bibinfo
  {volume} {08}},\ \bibinfo {pages} {036} (\bibinfo {year} {2013})},\ \Eprint
  {http://arxiv.org/abs/1210.7652} {arXiv:1210.7652 [astro-ph.CO]} \BibitemShut
  {NoStop}%
\bibitem [{\citenamefont {Keeley}\ and\ \citenamefont
  {Shafieloo}(2021)}]{Keeley:2021dmx}%
  \BibitemOpen
  \bibfield  {author} {\bibinfo {author} {\bibfnamefont {R.~E.}\ \bibnamefont
  {Keeley}}\ and\ \bibinfo {author} {\bibfnamefont {A.}~\bibnamefont
  {Shafieloo}},\ }\href@noop {} {\  (\bibinfo {year} {2021})},\ \Eprint
  {http://arxiv.org/abs/2111.04231} {arXiv:2111.04231 [astro-ph.CO]}
  \BibitemShut {NoStop}%
\bibitem [{\citenamefont {Koo}\ \emph {et~al.}(2021)\citenamefont {Koo},
  \citenamefont {Keeley}, \citenamefont {Shafieloo},\ and\ \citenamefont
  {L'Huillier}}]{Koo:2021suo}%
  \BibitemOpen
  \bibfield  {author} {\bibinfo {author} {\bibfnamefont {H.}~\bibnamefont
  {Koo}}, \bibinfo {author} {\bibfnamefont {R.~E.}\ \bibnamefont {Keeley}},
  \bibinfo {author} {\bibfnamefont {A.}~\bibnamefont {Shafieloo}}, \ and\
  \bibinfo {author} {\bibfnamefont {B.}~\bibnamefont {L'Huillier}},\
  }\href@noop {} {\  (\bibinfo {year} {2021})},\ \Eprint
  {http://arxiv.org/abs/2110.10977} {arXiv:2110.10977 [astro-ph.CO]}
  \BibitemShut {NoStop}%
\bibitem [{\citenamefont {Jeffreys}(1961)}]{Jeffreys61}%
  \BibitemOpen
  \bibfield  {author} {\bibinfo {author} {\bibfnamefont {H.}~\bibnamefont
  {Jeffreys}},\ }\href@noop {} {\emph {\bibinfo {title} {Theory of
  Probability}}},\ \bibinfo {edition} {3rd}\ ed.\ (\bibinfo  {publisher}
  {Oxford},\ \bibinfo {address} {Oxford, England},\ \bibinfo {year}
  {1961})\BibitemShut {NoStop}%
\bibitem [{\citenamefont {Kass}\ and\ \citenamefont
  {Raftery}(1995)}]{kass-raf}%
  \BibitemOpen
  \bibfield  {author} {\bibinfo {author} {\bibfnamefont {R.~E.}\ \bibnamefont
  {Kass}}\ and\ \bibinfo {author} {\bibfnamefont {A.~E.}\ \bibnamefont
  {Raftery}},\ }\href {\doibase 10.1080/01621459.1995.10476572} {\bibfield
  {journal} {\bibinfo  {journal} {Journal of the American Statistical
  Association}\ }\textbf {\bibinfo {volume} {90}},\ \bibinfo {pages} {773}
  (\bibinfo {year} {1995})},\ \Eprint
  {http://arxiv.org/abs/https://www.tandfonline.com/doi/pdf/10.1080/01621459.1995.10476572}
  {https://www.tandfonline.com/doi/pdf/10.1080/01621459.1995.10476572}
  \BibitemShut {NoStop}%
\bibitem [{\citenamefont {Farooq}\ \emph {et~al.}(2017)\citenamefont {Farooq},
  \citenamefont {Madiyar}, \citenamefont {Crandall},\ and\ \citenamefont
  {Ratra}}]{Farooq:2016zwm}%
  \BibitemOpen
  \bibfield  {author} {\bibinfo {author} {\bibfnamefont {O.}~\bibnamefont
  {Farooq}}, \bibinfo {author} {\bibfnamefont {F.~R.}\ \bibnamefont {Madiyar}},
  \bibinfo {author} {\bibfnamefont {S.}~\bibnamefont {Crandall}}, \ and\
  \bibinfo {author} {\bibfnamefont {B.}~\bibnamefont {Ratra}},\ }\href
  {\doibase 10.3847/1538-4357/835/1/26} {\bibfield  {journal} {\bibinfo
  {journal} {Astrophys. J.}\ }\textbf {\bibinfo {volume} {835}},\ \bibinfo
  {pages} {26} (\bibinfo {year} {2017})},\ \Eprint
  {http://arxiv.org/abs/1607.03537} {arXiv:1607.03537 [astro-ph.CO]}
  \BibitemShut {NoStop}%
%%CITATION = ARXIV:1607.03537;%%
\bibitem [{\citenamefont {Salvatier}\ \emph {et~al.}(2016)\citenamefont
  {Salvatier}, \citenamefont {Wiecki},\ and\ \citenamefont
  {Fonnesbeck}}]{pymc}%
  \BibitemOpen
  \bibfield  {author} {\bibinfo {author} {\bibfnamefont {J.}~\bibnamefont
  {Salvatier}}, \bibinfo {author} {\bibfnamefont {T.~V.}\ \bibnamefont
  {Wiecki}}, \ and\ \bibinfo {author} {\bibfnamefont {C.}~\bibnamefont
  {Fonnesbeck}},\ }\href@noop {} {\bibfield  {journal} {\bibinfo  {journal}
  {PeerJ Computer Science}\ }\textbf {\bibinfo {volume} {2}},\ \bibinfo {pages}
  {e55} (\bibinfo {year} {2016})}\BibitemShut {NoStop}%
\bibitem [{\citenamefont {Di~Valentino}\ \emph {et~al.}(2019)\citenamefont
  {Di~Valentino}, \citenamefont {Melchiorri},\ and\ \citenamefont
  {Silk}}]{DiValentino:2019qzk}%
  \BibitemOpen
  \bibfield  {author} {\bibinfo {author} {\bibfnamefont {E.}~\bibnamefont
  {Di~Valentino}}, \bibinfo {author} {\bibfnamefont {A.}~\bibnamefont
  {Melchiorri}}, \ and\ \bibinfo {author} {\bibfnamefont {J.}~\bibnamefont
  {Silk}},\ }\href {\doibase 10.1038/s41550-019-0906-9} {\bibfield  {journal}
  {\bibinfo  {journal} {Nature Astron.}\ }\textbf {\bibinfo {volume} {4}},\
  \bibinfo {pages} {196} (\bibinfo {year} {2019})},\ \Eprint
  {http://arxiv.org/abs/1911.02087} {arXiv:1911.02087 [astro-ph.CO]}
  \BibitemShut {NoStop}%
\end{thebibliography}%

\end{document}